\documentclass[twocolumn,english,showpacs,preprintnumbers,amsmath,amssymb]{revtex4}
\usepackage[T1]{fontenc}
\usepackage[latin9]{inputenc}
\usepackage{amsmath}
\usepackage{graphicx}
\usepackage{amssymb}

\makeatletter
\@ifundefined{textcolor}{}
{%
 \definecolor{BLACK}{gray}{0}
 \definecolor{WHITE}{gray}{1}
 \definecolor{RED}{rgb}{1,0,0}
 \definecolor{GREEN}{rgb}{0,1,0}
 \definecolor{BLUE}{rgb}{0,0,1}
 \definecolor{CYAN}{cmyk}{1,0,0,0}
 \definecolor{MAGENTA}{cmyk}{0,1,0,0}
 \definecolor{YELLOW}{cmyk}{0,0,1,0}
 }

\usepackage{dcolumn}% Align table columns on decimal point
\usepackage{bm}% bold math
\makeatother

\usepackage{babel}

\makeatother

\usepackage{babel}

\makeatother

\usepackage{babel}

\begin{document}

\title{Typicality in random matrix product states}

\author{Silvano Garnerone}

\email{garneron@usc.edu}

\affiliation{Department of Physics and Astronomy and Center for Quantum Information
Science \& Technology, University of Southern California, Los Angeles,
CA 90089 }

\author{Thiago R. de Oliveira}

\affiliation{Department of Physics and Astronomy and Center for Quantum Information
Science \& Technology, University of Southern California, Los Angeles,
CA 90089 }

\author{Paolo Zanardi}

\altaffiliation[Also at ]{Institute for Scientific Interchange, Viale Settimio Severo 65, I-10133 Torino, Italy}

\affiliation{Department of Physics and Astronomy and Center for Quantum Information
Science \& Technology, University of Southern California, Los Angeles,
CA 90089 }
\begin{abstract}
Recent results suggest that the use of ensembles in Statistical Mechanics
may not be necessary for isolated systems, since typically the states
of the Hilbert space would have properties similar to the ones of
the ensemble. Nevertheless, it is often argued that most of the states
of the Hilbert space are non-physical and not good descriptions of
realistic systems. Therefore, to better understand the actual power
of typicality it is important to ask if it is also a property of a
set of physically relevant states. Here we address this issue, studying
if and how typicality emerges in the set of matrix product states.
We show analytically that typicality occurs for the expectation value
of subsystems' observables when the rank of the matrix product state
scales polynomially with the size of the system with a power greater
than two. We illustrate this result numerically and present some indications
that typicality may appear already for a linear scaling of the rank
of the matrix product state. 
\end{abstract}

\pacs{Valid PACS appear here}

\maketitle

\section{introduction}

Statistical mechanics has been very successful in making predictions
about the behavior of macroscopic systems we encounter in nature,
yet there are still open questions concerning a fully quantum formulation
of statistical mechanics. For instance, how can we explain the use
of statistical ensembles in the description of a physical system which
is supposed to be in a definite state? In this context the independent
re-discovery by several groups of the importance of \textit{typicality}
\cite{GeMiMa,Leb,GoLeTu,PoShWi} has given rise to interesting directions
of research \cite{Rei1,Rei2,RiDuOl,Whi}. Although the concept originally
appeared as an incomplete formulation in a work by Schrodinger \cite{Sch},
Lebowitz was the one to coin the term 'typicality' and, together with
others, did some pioneering work on the subject \cite{Leb,GoLeTu}.
Typicality can be seen as a key feature justifying the effectiveness
of standard equilibrium statistical mechanics, without requiring ergodicity
or mixing. The works on typicality in the quantum setting have shown
that ensemble averages and subjective ignorance may not be necessary
concepts for the understanding of statistical mechanics \cite{PoShWi}.

Intuitively, typicality refers to the fact that the vast majority
of pure microstates of a quantum system, belonging to a well-defined
region of the allowed state space, yield measurement outcomes very
close to each other. More quantitatively, typicality can be associated
with a very small variance of the measurement outcomes with respect
to a specified ensemble of states.

Previous works \cite{PoShWi,Rei1,Rei2} have focused on the study
of typicality for \textit{general} quantum states, providing a first
alternative approach to the foundational problems of quantum statistical
mechanics. However it is well know that the generation of Haar distributed
random states is hard even at the quantum level \cite{EmWeSa}. Therefore
in order to consider typicality an effective scheme for the justification
of statistical mechanics one should restrict to realizable random
states, possibly with some specific physical content. First we need
to choose and characterize this set of states, though of course the
choice is not unique. In the present work we focus on Matrix Product
States (MPS) (see \cite{Verstraete08} for a review and original references)
as instances of physically accessible states. The reason why we restrict
the study of typicality to MPS is because these are a good example
of physically relevant states, arising as ground states of local Hamiltonians
and being at the basis of some of the most recent and promising classical
algorithms for the simulation of quantum systems \cite{Verstraete08}.
Both of these properties justify a better understanding of their statistical
properties with respect to typicality, which eventually can also lead
to new powerful simulation techniques (as the work in \cite{Whi}
may suggest). We shall prove that typicality can emerge in the MPS
set, and then illustrate this result with some numerical simulations.

\section{random matrix product states}

A matrix product state is a pure one-dimensional quantum state whose
coefficients are specified by a product of matrices. In the case of
Periodic Boundary Conditions (PBC) an MPS can be written as \[
|\psi\rangle=\sum_{i_{1},...,i_{N}}{\rm Tr}\left[A^{i_{1}}[1]\cdots A^{i_{N}}[N]\right]|i_{1}\cdots i_{N}\rangle,\]
 whereas for Open Boundary Conditions (OBC) we have \[
|\psi\rangle=\sum_{i_{1},...,i_{N}}\langle\phi_{I}|A^{i_{1}}[1]\cdots A^{i_{N}}[N]|\phi_{F}\rangle|i_{1}\cdots i_{N}\rangle,\]
 with $|\phi_{I}\rangle$ and $|\phi_{F}\rangle$ specifying the states
at the boundaries and $|i_{k}\rangle$ a local basis at site $k$.
The matrices $\{A^{1}[s],A^{2}[s],\dots,A^{D}[s]\}$, with $s\in\{1,\dots,N\}$,
are $\chi$-dimensional complex matrices, with $D$ the local Hilbert
space dimension. For homogeneous MPS the set $\{A^{1}[s],A^{2}[s],\dots,A^{D}[s]\}$
is the same for all sites $s$. In the case of PBC they are referred
to as Translationally Ivariant (TI) MPS. In the present work, for
simplicity of notation and analysis, we deal numerically with OBC-MPS
and analytically with PBC-TI-MPS. We checked numerically that all
of our conclusions hold true independently of the boundary conditions
and invariance under translations.

By definition, a MPS is specified by the set $\{A^{1},A^{2},\dots,A^{D}\}$,
though there may exist a different set of matrices that form the same
MPS. In \cite{PeVeWo} it is shown that this sort of gauge degree
of freedom can be fixed using a canonical form where the $A$ matrices
satisfy two constraints: $\sum_{i=1}^{D}A^{i}A^{i\dagger}=\mathbb{I}$
and $\sum_{i=1}^{D}A^{i\dagger}\Lambda A^{i}=\Lambda$, for fixed
$\Lambda$ (an alternative set of constraints is given by $\sum_{i=1}^{D}A^{i\dagger}A^{i}=\mathbb{I}$
and $\sum_{i=1}^{D}A^{i}\Lambda A^{i\dagger}=\Lambda$, see \cite{PeVeWo}
for details). MPS can also be seen as generalized valence-bond states
\cite{Verstraete08}, and as such, emerging from the projection on
some virtual or ancillary Hilbert space. The fundamental parameter
of an MPS is the size $\chi$ of the $A$-matrices. In general, an
MPS contains $ND\chi^{2}$ parameters, much less than the usual $D^{N}$
of a general state, and as a consequence the maximum entanglement
a subsystem can have with its environment depends on $\chi$. It can
be shown that any state can be described as an MPS for large enough
$\chi$ with at most $\chi\propto D^{N}$ (though there is no advantage
in such a description) \cite{Vid}. For more details and properties
of MPS used in this work see appendix B.

\begin{figure}
\includegraphics[scale=0.4]{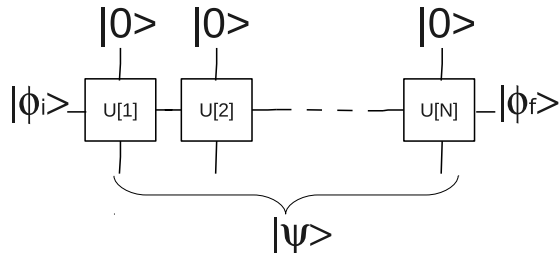}

\caption{Sequential generation of an MPS $|\psi\rangle,$ with $|\phi_{i}\rangle$
and $|\phi_{f}\rangle$ the boundary normalized states. }

\label{fig:fig1}
\end{figure}

For our purposes we need to generate an ensemble of Random MPS (RMPS)
and the way to do this is by no means unique. One could think, for
example, of choosing a set of matrices $\{A^{1},\dots,A^{D}\}$ belonging
to some relevant ensemble known in random matrix theory. This choice
would induce additional symmetries on the $A$-matrices that will
constrain the set of RMPS too much, and for which the physical meaning
would not be clear a priori (see \cite{Has} for a related construction
in a different context). The ensemble of RMPS that we consider in
this work is constructed by the repeated random unitary interaction
between an ancilla and a physical system, as described in the framework
of the sequential generation of MPS \cite{ScHaWo,PeVeWo}. This is
an operationally and physically motivated realization of MPS. We now
briefly summarize the construction in \cite{ScHaWo}. Consider a spin
chain initially in a product state $|0\rangle^{\otimes N}\in\mathcal{H}_{B}^{\otimes N}$
(with $\mathcal{H}_{B}\simeq\mathbb{C}^{D}$) and an ancillary system
in the state $|\phi_{I}\rangle\in\mathcal{H}_{A}\simeq\mathbb{C}^{\chi}$.
Let $U[k]$ be a unitary operation on $\mathcal{H}_{A}\otimes\mathcal{H}_{B}$,
acting on the ancillary system and the $k$'th site of the chain (see
Fig.\ref{fig:fig1}). The $A[k]$ matrices are defined by \begin{equation}
A_{\alpha,\beta}^{i}[k]\equiv\langle i,\alpha|U[k]|\beta,0\rangle,\label{eq:Amat}\end{equation}
 where the greek indeces refer to the ancilla space and the latin
indeces to the physical space. For homogeneous MPS the unitary interaction
is the same for all the sites in the spin chain. Due to unitarity
we have $\sum_{i}A^{i}[k]^{\dagger}A^{i}[k]=\mathbb{I}_{\chi}$ for
all $k$ in the bulk. This property, together with a proper normalization
of the boundaries, corresponds to an MPS of unit-norm (see appendix
B for more details). Letting the ancilla interact sequentially with
the $N$ sites of the chain and assuming that the ancilla decouples
in the last step (this can be done without loss of generality, as
shown in \cite{ScHaWo}), the state on $\mathcal{H}_{B}^{\otimes N}$
is described by \[
|\psi\rangle=\sum_{i_{1},\dots,i_{N}}\langle\phi_{F}|A^{i_{N}}\cdots A^{i_{1}}|\phi_{I}\rangle|i_{N}\cdots i_{1}\rangle,\]
which is a homogeneous MPS with OBC. It can be proved \cite{ScHaWo,PeVeWo}
that the set of states generated in this way is equal to the set of
OBC-MPS. We choose the interaction characterizing the homogeneous
RMPS ensemble to be represented by a random unitary matrix $U$ distributed
according to the Haar measure.

Since any state can be described by an MPS when $\chi\propto D^{N}$
\cite{Vid}, typicality should appear trivially for MPS with this
exponential scaling of $\chi$ in $N$. Therefore the relevant question
is if it is possible to have typicality when the rank increases at
most polynomially with the number of particles: $\chi={\rm \mathit{poly}(N)}$.
This will be the subject of the next section.

\section{typicality in rmps}

Typicality can be studied at a more formal level in the framework
of concentration of measure, a mathematical tool which allows to establish
typicality in large-dimensional Hilbert spaces \cite{MiSc}. The concentration
of measure phenomenon allows to quantify the probability of fluctuations
for functions of random variables, and in the physical literature
has already been applied in a variety of contexts \cite{HaLeSh,BrMoWi,GrFlEi,Low}.
We shall use a result on the concentration of measure phenomenon for
the unitary group in order to prove typicality for subsystems' observables
(a more rigorous mathematical introduction to the topic can be found
in \cite{MiSc}; in particular, here we use theorem 6.7.1 of that
book). Concentration of measure holds for the unitary group and this
means that there exist universal positive constants $c_{1}$ and $c_{2}$
such that for any function $f:U(d)\rightarrow\mathbb{R},$ from the
set of Haar-distributed unitary matrices of size $d\times d$ into
$\mathbb{R}$, and with Lipschitz constant $\eta$ \begin{equation}
\textrm{Pr}\left[\left|f-\bar{f}\right|\geq\epsilon\right]\leq c_{1}\exp\left(-c_{2}\epsilon^{2}d/\eta^{2}\right),\label{Eq:Prob}\end{equation}
 where $\bar{f}$ denotes the average value of $f$. From this expression
one can see that typicality is valid only for functions $f$ for which
the ratio $d/\eta^{2}$ (where $\eta$ can in principle also depend
on $d$) increases with the dimension $d$ of the domain, since in
this case the probability of large fluctuations around their average
will decrease exponentially in $d$.

In the present work the random variable $f$ will be the expectation
value of an observable with respect to a random MPS. The observables
that we consider are those that can be expressed as the tensor product
of local observables. In the usual transfer matrix notation for normalized
MPS (see appendix B) we can write \[
f\equiv\textrm{Tr}\left[\prod_{k=1}^{N}E_{O[k]}\right]\]
 with \[
E_{O[k]}\equiv\sum_{i_{k},j_{k}=1}^{D}\langle i_{k}|O[k]|j_{k}\rangle A^{i_{k}}[k]\otimes A^{j_{k}}[k]^{*}\]
 the tranfer matrix associated to the observable $O\equiv\bigotimes_{k=1}^{N}O[k]$.
The $A[k]$-matrices characterizing the state are obtained as sub-blocks
of random unitaries $U[k]$, analogusly to Eq.(\ref{eq:Amat}). In
this way the expectation value of the tensor product of local observables
can be seen as a random variable $f:U(\chi D)\rightarrow\mathbb{R}$,
from the set of uniformly distributed unitary matrices of size $\chi D\times\chi D$
into $\mathbb{R}.$ In order to apply the concentration of measure
result for functions of random unitaries we need to find an upper-bound
for the Lipschitz constant $\eta$ in (\ref{Eq:Prob}) \[
\eta\equiv sup_{U_{1}\neq U_{2}}\frac{|f(U_{1})-f(U_{2})|}{\Vert U_{1}-U_{2}\Vert_{2}}.\]
 In order to do that we look for an upper-bound of the absolute value
of the differential of $f$ \[
|df|=|dTr\left[\prod_{k=1}^{N}E_{O[k]}\right]|.\]
 We consider the case of subsystems of size $L$ specified by observables
$O$ of this form \[
O\equiv\left(\bigotimes_{k=1}^{L}O[k]\right)\left(\bigotimes_{k=L+1}^{N}\mathbb{I}[k]\right).\]
 We will refer to the rest of the chain, of $N-L$ sites, as the environment
or bath. Using standard properties of the differential calculus for
matrices we have (see appendix A) \[
|df|=\left|dTr[\prod_{k=1}^{L}E_{O[k]}E_{\mathbb{I}}^{N-L}]\right|\leq\left|{\rm Tr}\left[\left(d\prod_{k=1}^{L}E_{O[k]}\right)E_{\mathbb{I}}^{N-L}\right]\right|\]
 \begin{equation}
+\left|{\rm Tr}\left[\prod_{k=1}^{L}E_{O[k]}dE_{\mathbb{I}}^{N-L}\right]\right|.\label{eq:firstbound}\end{equation}
 Let us consider the first term on the right-hand side in (\ref{eq:firstbound}).
We have that \[
\left|{\rm Tr}\left[\left(d\prod_{k=1}^{L}E_{O[k]}\right)E_{\mathbb{I}}^{N-L}\right]\right|\]
 \[
\leq\sum_{k=1}^{L}\prod_{j\neq k}^{L}\Vert E_{O[j]}\Vert_{\infty}\Vert dE_{O[k]}\Vert_{\infty}\Vert E_{\mathbb{I}}^{N-L}\Vert_{1},\]
 where $\Vert\cdot\Vert_{k}$ stands for the matrix $k$-norm. We
denote with $\Vert O\Vert_{\infty}$ the $max\left\{ \Vert O[k]\Vert_{\infty}\,;\; k=1,\dots,L\right\} $
and use the following relations \[
\Vert A^{i}\Vert_{\infty}=\Vert(|i\rangle\langle i|_{B}\otimes\mathbb{I}_{A})U(|0\rangle\langle0|_{B}\otimes\mathbb{I}_{A})\Vert_{\infty}\leq\Vert U\Vert_{\infty}=1,\]
 \[
\Vert E_{\mathbb{I}}^{N-L}\Vert_{1}\leq1+\chi^{2}\;\epsilon_{2}^{N-L},\]
 \[
\Vert dE_{O[j]}\Vert_{\infty}\leq2D^{2}\Vert O\Vert_{\infty}\Vert dU\Vert_{\infty},\]
 where $\epsilon_{2}$ is the second-largest eigenvalue in the spectrum
of the transfer matrix over the ensemble of RMPS (remember that for
normalized MPS the largest eigenvalue of $E_{\mathbb{I}}$ is $1$,
so that $\epsilon_{2}<1$ for all the realizations). We can then bound
the first term of (\ref{eq:firstbound}) with \begin{equation}
2LD^{2L}\left(1+\chi^{2}\;\epsilon_{2}^{N-L}\right)\Vert O\Vert_{\infty}^{L}\Vert dU\Vert_{\infty}.\label{eq.bound1}\end{equation}
 Let us now consider the second term in the right hand side of (\ref{eq:firstbound}),
which can be written as \[
\left|\sum_{j=L+1}^{N}{\rm Tr}\left[\prod_{k=1}^{L}E_{O[k]}E_{\mathbb{I}}^{j-1-L}dE_{\mathbb{I}[j]}E_{\mathbb{I}}^{N-j}\right]\right|.\]
 We can bound each term in this sum with \[
\Vert\prod_{k=1}^{L}E_{O[k]}E_{\mathbb{I}}^{\lfloor\frac{N-L}{2}\rfloor}dE_{\mathbb{I}[j]}\Vert_{\infty}\Vert E_{\mathbb{I}}^{\lfloor\frac{N-L}{2}\rfloor}\Vert_{1},\]
 and the total sum with \[
\left(1+\chi^{2}\;\epsilon_{2}^{\lfloor\frac{N-L}{2}\rfloor}\right)\left(N-L-1\right)\Vert dE_{\mathbb{I}}\Vert_{\infty}\Vert\prod_{k=1}^{L}E_{O[k]}\Vert_{\infty},\]
 which is smaller than or equal to \begin{equation}
2\left(N-L-1\right)D^{2L+2}\left(1+\chi^{2}\;\epsilon_{2}^{\lfloor\frac{N-L}{2}\rfloor}\right)\Vert O\Vert_{\infty}^{L}\Vert dU\Vert_{\infty}.\label{eq.bound2}\end{equation}
 The total variation of the expectation value is then bounded by the
sum of equation (\ref{eq.bound1}) and equation (\ref{eq.bound2}),
which in the relevant regime of interest ($N\gg L$) becomes \[
4D^{2L+2}N\Vert O\Vert_{\infty}^{L}\Vert dU\Vert_{2},\]
 where we used the fact that $\Vert\cdot\Vert_{\infty}\leq\Vert\cdot\Vert_{2}$.
The Lipschitz constant $\eta$ for the function $f$ is then upper-bounded
by \[
\eta\leq4D^{2L+2}N\Vert O\Vert_{\infty}^{L},\]
for $N\gg L$. Along with Eq.(\ref{Eq:Prob}), this implies that increasing
the size of the environment will cause the expectation value of the
observables of any subsystem to concentrate, provided $\chi(N)$ increases
faster than $N^{2}$ \[
\textrm{Pr}\left[\left|f-\bar{f}\right|\geq\epsilon\right]\leq c_{1}\exp\left(-c_{2}^{\prime}\epsilon^{2}\chi(N)/N^{2}\right),\]
 where we have absorbed all the constant in $c_{2}^{\prime}$. It
is important to notice that the set of MPS, for fixed $\chi$ and
$N$, is exponentially small with respect to the total number of states
in the same Hilbert space. As $N$ increases the dimension of the
Hilbert space will increase exponentially but a polynomial scaling
of $\chi$ in $N$ will be sufficient to guarantee typicality. This
shows that typicality is a property of a class of accessible states
of quantum system, extending some of the implications of previous
work \cite{PoShWi,Rei1,Rei2} to an experimentally and computationally
accessible regime. %
\begin{figure}[htp]
\centering \includegraphics[scale=0.65]{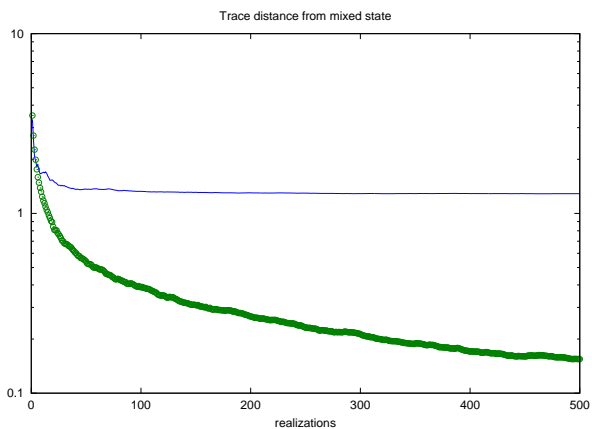} \caption{Trace distance between the average OBC-MPS with $\chi=2$  and the
completely mixed state. The lower line is the trace distance between
the average uniformely distributed random states and the completely
mixed state.}

\label{fig:fig2} 
\end{figure}

\begin{figure}[htp]
\centering \includegraphics[scale=0.65]{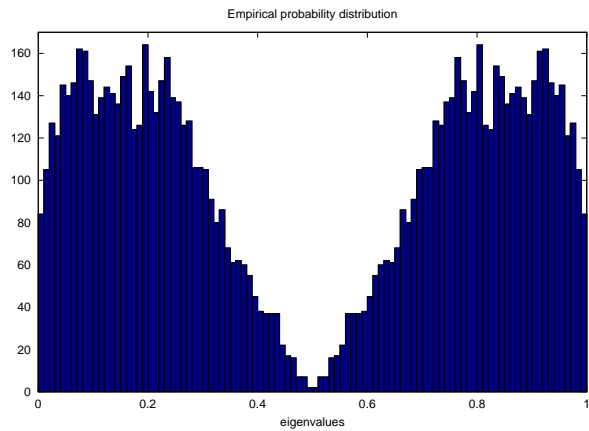} \caption{Empirical probability distribution for the eigenvalues of the reduced
density matrix of RMPS (5000 states). The subsystem has dimension
2 and the total system has dimension 16 ($\chi=2$).}

\label{fig:fig3} 
\end{figure}

\section{numerical results}

\begin{figure}
\includegraphics[scale=0.65]{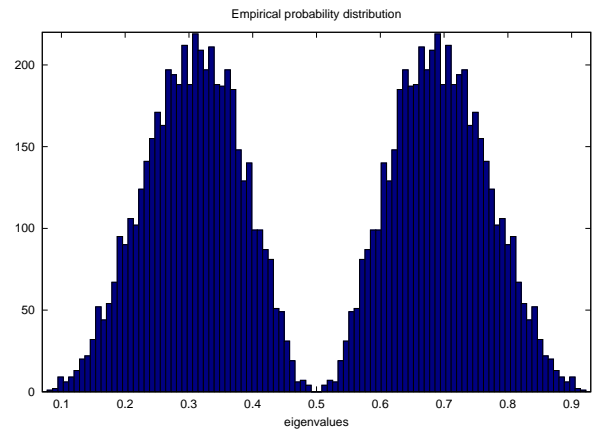}

\caption{Empirical probability distribution for the eigenvalues of the reduced
density matrix of a general Haar-distributed random state (5000 realizations).
The subsystem has dimension 2 and the total system has dimension 16.}

\label{fig:fig4}
\end{figure}

\begin{figure}
\includegraphics[scale=0.3]{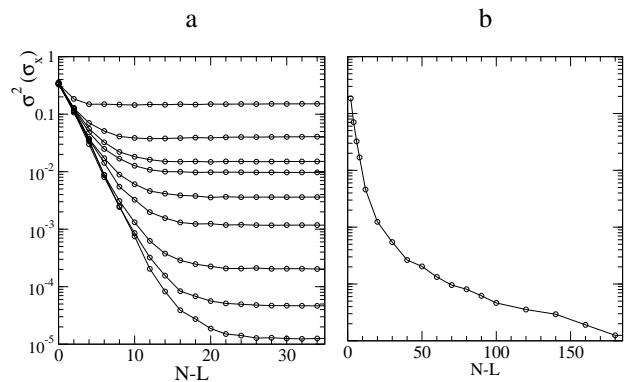}

\caption{(a) The variance of the expectation value of $\sigma_{x}$ ($L=1$)
increasing the size of the system and for fixed but different values
of $\chi=2,4,6,8,12,20,50,100,\;\text{and}\;180$ (from top to bottom).
(b) The variance of the expectation value of $\sigma_{x}$ ($L=1$)
for increasing system size when the MPS dimension increases linearly
with the number of particles in the bath: $\chi=N-L$. }

\label{fig:fig5}
\end{figure}

The recent new approach to typicality comes in two main flavors, one
due to Popescu et. al. \cite{PoShWi} and a second one due to Reimann
\cite{Rei1}. The first approach is more mathematical in two aspects:
it uses results from the concentration of measure phenomenon and considers
distances between states of subsystems. Reimann, on the other hand,
uses more heuristic arguments and studies general observables. In
this latter approach one has to show that the variance of the expectation
values of observables is small and decreasing with the size of the
bath. This implies, by the Chebyshev inequality, a concentration of
measure result. With the approach of \cite{PoShWi} one can directly
study the fluctuations of the trace distance between the states in
the ensemble and their average. A concentration result with this approach
is a stronger result, in the sense that it is sufficient for having
typicality at the level of observables. On the other hand typicality
for all local observables can also imply a weaker concentration result
for the state of the subsystem (see appendix C). In the numerical
simulations we considered both the variance of the expectation value
of observables and the fluctuations in the subsystem of the distance
of RMPS from their average state $\overline{\rho}\equiv\overline{|\psi\rangle\langle\psi|}$,
for which an exact expression can be obtained and reads in terms of
the components \[
\overline{\rho_{{\bf \mathbf{i}},{\bf \mathbf{j}}}}={\rm Tr}_{\mathcal{H}_{A\otimes B}^{\otimes2N}}\left[\left(|{\bf \mathbf{0}},{\bf \mathbf{j}}\rangle\langle{\bf \mathbf{i}},{\bf \mathbf{0}}|\otimes T\otimes T^{\dagger}\right)\overline{U^{\otimes N}\otimes U^{\dagger\otimes N}}\right],\]
 where the $N-$components vectors are defined by ${\bf \mathbf{0}}\equiv(0,...,0),$
${\bf \mathbf{i}}\equiv(i_{1},\dots,i_{N})$ and the same for ${\bf \mathbf{j}}.$
The operator $T$ acts on the $N-$tensor product of the ancilla system
and cyclicly permutes the components \[
T|\alpha\rangle_{1}\dots|\alpha\rangle_{N}=|\alpha\rangle_{N}|\alpha\rangle_{1}\dots|\alpha\rangle_{N-1}.\]
 The unitary random matrix $U(\chi D)$ is the one used in the sequential
generation of the random MPS. A closed form for the average of the
tensor product of unitaries $\overline{U^{\otimes N}\otimes U^{\dagger\otimes N}}$
is known \cite{CoSn}. In Fig.\ref{fig:fig2} we plot the trace distance
between the average random MPS with OBC and the completely mixed state
(for a chain of $4$ qubits) $\Vert\overline{\rho}-\frac{\mathbb{I}}{16}\Vert_{1}$,
as a function of the size of the sampling set (the number of randomly
generated states). In the same figure we also plot the same quantity
in the case of random general pure state (not necessarily MPS). As
can be seen from the figure the average OBC-RMPS is at a finite distance
from the mixed state, while the average general state approaches the
mixed state increasing the number of sampled states. Another distinctive
feature of the homogeneous OBC-RMPS is shown in Fig.\ref{fig:fig3},
where we plot the empirical probability distribution of the eigenvalues
of the reduced density matrix of 1 qubit in a 4-sites RMPS. In Fig.\ref{fig:fig4}
we show the same plot for general randomly generated states. As can
be seen the two distribution differs significantly. From now on, unless
otherwise stated, for all simulations we consider an ensemble of $500$
RMPS which originate from random unitaries distributed according to
the Haar measure. We now want to illustrate our analytical results
studying the behavior of the variance of $\sigma_{x}$, which acts
on a particle ($L=1$) in the middle of the chain. Note that the important
variable is not the absolute size of the subsystem or the bath but
the ratio between them. As can be seen in Fig.\ref{fig:fig5}.a, when
we fix the value of $\chi$ and increase the number of qubits in the
bath the variance starts to decrease, but soon reaches a limiting
value and does not decrease any more. The limiting value depends on
$\chi$, becoming smaller as $\chi$ is increased. This could be expected
from our bound and from a known property of MPS: correlations between
system and environment are of finite range and depend on $\chi$.
This is also consisten with recent result on finite entanglement scaling
at criticality \cite{PoMuTu}. Note, however, that our analytical
result does not exclude the possibility of having typicality for fixed
$\chi$ or $\chi$ scaling linearly with $N$. It only guarantees
typicality in the case of a scaling greater than quadratic. We then
analyze the case where $\chi=N-L$, as show in Fig.\ref{fig:fig5}.b.
There it can be seen that until $\chi=N-L=180$ the variance is decreasing
monotonically, which indicates that typicality can emerge already
for a linear scaling of $\chi$ with the number of particles. However,
at the present moment our simulations do not allow for a conclusive
statement about the precise scaling of $\chi$ with $N-L$ that assures
typicality.

\begin{figure}
\includegraphics[scale=0.3]{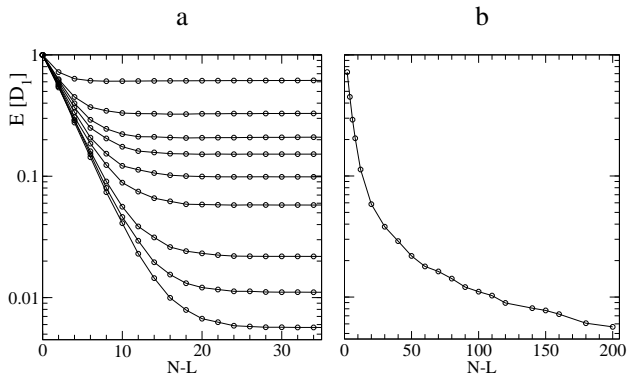}

\caption{(a) The average trace distance from the average state of the ensemble
of RMPS as the number of qubits in the environment ($N-L$) increases,
for fixed but different values of $\chi=2,4,6,8,12,20,50,100,\;\text{and}\;200$
(from top to bottom) and for $L=1$. (b) The average trace distance
from the average state of the ensemble of RMPS as the number of qubits
in the environment ($N-L$) increases, taking the MPS dimension to
increase linearly with the number of particles in the bath: $\chi=N-L$.}

\label{fig:fig6}
\end{figure}

We also investigate the behavior of the average trace distance from
the average state at the level of the sub-system, which is one particle
in our case. Denoting with $\rho_{s}$ the reduced density matrix
of the subsystem of a RMPS and $\overline{\rho_{s}}$ the average
MPS, Fig.(\ref{fig:fig6}.a) shows the dependence on the size of the
bath of the average value $E(D_{1})$ of the trace distance $D_{1}\equiv\Vert\rho_{s}-\overline{\rho_{s}}\Vert_{1}$.
In general, we expect that having typicality at the level of states
is harder than at the level of observables (see appendix C). Again
we look at the case of fixed $\chi$ (Fig.(\ref{fig:fig6}.a)) and
$\chi=N-L$ (Fig.(\ref{fig:fig6}.b)). The conclusions are similar
to the case of observables, and it appears that even at the level
of states typicality may occur already for a linear scaling of $\chi$
in the size of the bath, while the behavior for different but fixed
values of $\chi$ is again consistent with well-known MPS properties.

\section{conclusions}

In summary, we have shown that typicality can arise not only for an
exponentially big Hilbert space but also for a physically accessible
smaller set of states: the Matrix Product States. More specifically,
we showed analytically that typicality occurs for MPSs having $\chi$
scaling polynomially in the size of the system (with a power greater
than 2). We then presented some numerical calculations which indicate
that typicality may already emerge for a linear scaling of $\chi$
in the system size at the level of both observables and the states.

Our results provide further evidence that typicality may play a role
in a better understanding of the foundations of statistical mechanics.
Nonetheless, there are still some aspects that require a deeper analysis.
For example, it would be interesting to have more information on the
average state obtained from the present ensemble of RMPS (in general
$\overline{\rho}$ will still be a matrix product state but with a
bigger rank than its components) and to better characterize the role
of the geometry of the partition in between subsystem and bath and
their correlations. 

\smallskip{}

\begin{acknowledgments}
We would like to thank L. Campos-Venuti, G. Chiribella, D.Perez-Garcia
for useful discussions and N. T. Jacobson for a careful reading of
the manuscript. We thank J.I. Cirac for suggesting the use of the
construction in \cite{ScHaWo} and D. Gross for pointing us to \cite{MiSc}.
We thank also the Benasque Center for Science where part of the work
was done in a very informal and fruitful atmosphere. Supported by
NSF grants: PHY-803304,DMR-0804914.\end{acknowledgments}

\appendix
%dummy comment inserted by tex2lyx to ensure that this paragraph is not empty
\bigskip{}

\textbf{APPENDIX A. }If $f$ is a real-valued function on a metric
space $(X,d)$, its Lipschitz constant is \[
\vert f\vert_{\mathcal{L}}=sup_{x\neq y}\frac{\vert f(x)-f(y)\vert}{d(x,y)},\]
 for $x,y\in X$. If $X=\mathbb{R}^{n},$ we let $d(x,y)=\Vert x-y\Vert,$
with \[
\Vert x\Vert=\left(\sum_{i=1}^{N}x_{i}^{2}\right)^{\frac{1}{2}}\]
 the euclidean norm. The above notation can be easily applied also
to the case when $X=\mathcal{M}_{n}(\mathbb{C})$ is the set of $n\times n$
matrices. For $A\in\mathcal{M}_{n}(\mathbb{C})$ we define \[
\Vert A\Vert_{\infty}=sup_{|\psi\rangle}\frac{\Vert A|\psi\rangle\Vert}{\Vert|\psi\rangle\Vert},\]
\[
\Vert A\Vert_{2}=\sqrt{Tr(A^{\dagger}A)},\]
\[
\Vert A\Vert_{1}=Tr|A|.\]
All this norms are unitarily invariant and submultiplicative. In this
work we used also the following relation \[
Tr(AB)\leq\Vert A\Vert_{\infty}\Vert B\Vert_{1}.\]

In the derivation of the upper bound for the Lipschitz constant we
made use of standard properties of differentiation with respect to
a matrix (\cite{MaNe}) \[
dTr(X)=Tr(dX),\]
\[
d(XY)=(dX)Y+X(dY),\]
where $X$ and $Y$ are arbitrary matrices. Defining the differential
$df(X)$ to be the part of $f(X+dX)-f(X)$ which is linear in $dX$,
the gradient $\nabla f$ satisfies $df=\nabla f\cdot dX$. Since the
Lipschitz constant is equivalent to the $sup|\nabla f|,$ an upper-bound
for the gradient will provide an upper-bound for the Lipschitz constant.

\smallskip{}

\textbf{APPENDIX B}. Here we review some of the notation used in the
literature on MPS. A detailed exposition can be found in \cite{Verstraete08}.
For brevity we will focus for the moment on normalized MPS with periodic
boundary conditions, which by definition is a state that can be written
as\[
|\psi\rangle=\sum_{i_{1},\dots,i_{N}=1}^{D}Tr(A^{i_{1}}[1]A^{i_{2}}[2]\dots A^{i_{N}}[N])|i_{1}i_{2}\dots i_{N}\rangle,\]
 where the matrices $A$ are $\chi\times\chi$ complex matrices, labeled
by the site index $\in\{1,\dots,N\}$ and by the local bases index
$\in\{1,\dots,D\}.$

The expectation value of some operator $S,$ which is the tensor product
of local operators $S[k]$ at each site $k$ \[
S=S[1]\otimes\dots\otimes S[N],\]
is given by $\langle\psi|\bigotimes_{k=1}^{N}S[k]|\psi\rangle$ which
is equal to\[
\sum_{i_{1},i_{1}^{\prime}\dots,i_{N},i_{N}^{\prime}}^{D}Tr\left(\prod_{k=1}^{N}A^{i_{k}}[k]\right)Tr\left(\prod_{k=1}^{N}A^{i_{k}^{\prime}}[k]^{*}\right)\]
\[
\times\prod_{k=1}^{N}\langle i_{k}^{\prime}|S[k]|i_{k}\rangle\]
\[
=\sum_{i_{1},i_{1}^{\prime}\dots,i_{N},i_{N}^{\prime}}^{D}Tr\left(\prod_{k=1}^{N}A^{i_{k}}[k]\otimes\prod_{k=1}^{N}A^{i_{k}^{\prime}}[k]^{*}\right)\]
\[
\times\prod_{k=1}^{N}\langle i_{k}^{\prime}|S[k]|i_{k}\rangle\]
\[
=\sum_{i_{1},i_{1}^{\prime}\dots,i_{N},i_{N}^{\prime}}^{D}Tr\left[\prod_{k=1}^{N}\left(A^{i_{k}}[k]\otimes A^{i_{k}^{\prime}}[k]^{*}\right)\right]\prod_{k=1}^{N}\langle i_{k}^{\prime}|S[k]|i_{k}\rangle\]
\[
=Tr\left[\sum_{i_{1},i_{1}^{\prime}\dots,i_{N},i_{N}^{\prime}}^{D}\prod_{k=1}^{N}\langle i_{k}^{\prime}|S[k]|i_{k}\rangle\left(A^{i_{k}}[k]\otimes A^{i_{k}^{\prime}}[k]^{*}\right)\right]\]
\[
=Tr\left[\prod_{k=1}^{N}\sum_{i_{k},i_{k}^{\prime}}^{D}\langle i_{k}^{\prime}|S[k]|i_{k}\rangle\left(A^{i_{k}}[k]\otimes A^{i_{k}^{\prime}}[k]^{*}\right)\right].\]
 Defining the transfer matrix or transfer operator \[
E_{S[k]}[k]\equiv\sum_{i_{k},i_{k}^{\prime}}^{D}\langle i_{k}^{\prime}|S[k]|i_{k}\rangle\left(A^{i_{k}}[k]\otimes A^{i_{k}^{\prime}}[k]^{*}\right),\]
we see that \[
\langle\psi|\bigotimes_{k=1}^{N}S[k]|\psi\rangle=Tr\left[\prod_{k=1}^{N}E_{S[k]}[k]\right].\]
 The normalization of the state is given by \begin{equation}
Tr(\prod_{k=1}^{N}E_{\mathbb{I}[k]}[k]),\label{eq:norm}\end{equation}
 with $S[k]=\mathbb{I}[k]$ for all $k.$

Let us now consider the case of sequentially generated OBC-MPS \cite{ScHaWo}.
Consider a spin chain initially in a product state \[
|0\rangle^{\otimes N}\in\mathcal{H}_{B}^{\otimes N},\]
 and an ancillary system initially in the state $|\phi_{I}\rangle\in\mathcal{H}_{A}.$
Let us introduce a unitary operator $U[k]$ acting on $\mathcal{H}_{A}\otimes\mathcal{H}_{B}$,
for each site $k$ in the chain. Defining \[
A_{\alpha,\beta}^{i}[k]\equiv\langle i,\alpha|U[k]|\beta,0\rangle,\]
 unitarity implies the following\begin{equation}
U[k]^{\dagger}U[k]=\mathbb{I}_{D\chi}\implies\sum_{i}A^{i}[k]^{\dagger}A^{i}[k]=\mathbb{I}_{\chi}\quad k\in\{1,\dots,N\}.\label{eq:unitarity}\end{equation}
 Letting the ancilla interact sequentially with all the sites in the
chain (see Fig.\ref{fig:fig1}) we obtain the state\textbf{\[
|\psi\rangle=\sum_{i_{1},\dots,i_{N}=1}^{D}\langle\phi_{F}|A^{i_{N}}[1]\dots A^{i_{1}}[N]|\phi_{I}\rangle|i_{N}\dots i_{1}\rangle.\]
}Let us write the OBC-MPS in a different way ($A^{i_{N+1}}[N+1]\equiv\langle\phi_{F}|$
and $A^{i_{0}}[0]\equiv|\phi_{I}\rangle$)\[
|\psi\rangle=\sum_{i_{1},\dots,i_{N}=1}^{D}A^{i_{N+1}}[N+1]A^{i_{N}}[1]\times\]
\[
\times A^{i_{1}}[N]A^{i_{0}}[0]|i_{N+1}i_{N}\dots i_{1}i_{0}\rangle,\]
where $A^{i_{k}}[k]$ are $D_{k+1}\times D_{k}$ matrices with $D_{0}=D_{N+1}=1$
and $D_{k}=\chi$ with $k\in\{1,\dots,N\}$. The normalization of
the MPS is given by\[
\langle\psi|\psi\rangle=\sum_{i_{0},\dots,i_{N+1}=1}^{D}A^{i_{0}}[0]^{\dagger}A^{i_{1}}[1]^{\dagger}\dots A^{i_{N+1}}[N+1]^{\dagger}\times\]
\[
\times A^{i_{N+1}}[N+1]A^{i_{N}}[N]\dots A^{i_{1}}[1]A^{i_{0}}[0].\]
Using a singular value decomposition \cite{PeVeWo} it is always possible
to find a canonical form for $|\psi\rangle$ such that\[
\sum_{i_{N+1}}A^{i_{N+1}}[N+1]^{\dagger}A^{i_{N+1}}[N+1]=\mathbb{I}_{\chi}.\]
Using (\ref{eq:unitarity}) recursively one has \[
\langle\psi|\psi\rangle=\sum_{i_{0}}A^{i_{0}}[0]^{\dagger}A^{i_{0}}[0]=1,\]
 where the last equality follows from a normalization condition which
can be imposed without loss of generality on the boundary local matrix
$A^{i}[0]$ \cite{PeVeWo}.

In the case of periodic boundary conditions the norm of the state
\[
|\psi\rangle=\sum_{i_{1},\dots,i_{N}=1}^{D}Tr(A^{i_{1}}[1]A^{i_{2}}[2]\dots A^{i_{N}}[N])|i_{1}i_{2}\dots i_{N}\rangle,\]
is given by Eq.\ref{eq:norm} with $E_{\mathbb{I}}[k]\equiv\sum_{i_{k}=1}^{D}A^{i_{k}}[k]\otimes A^{i_{k}}[k]^{*}.$
For simplicity of notation, but without loss of generality, we shall
restrict now to the translational invariant case. To any MPS it can
always be associated a Completely Positive (CP) map \cite{PeVeWo}
\[
\mathcal{E}(X)\equiv\sum_{i}A_{i}XA_{i}^{\dagger}.\]
 The CP map can always be assumed to have spectral radius 1 (corresponding
to the absolute value of its maximum singular value) \cite{PeVeWo}.
The map $\mathcal{E}$ and $E_{\mathbb{I}}$ have the same spectrum
since \[
\langle\beta_{1}|\mathcal{E}\left(|\alpha_{1}\rangle\langle\alpha_{2}|\right)|\beta_{2}\rangle=\langle\beta_{1},\beta_{2}|E_{\mathbb{I}}|\alpha_{1},\alpha_{2}\rangle.\]
 Assuming that $\mathcal{E}$ has only one eigenvalue equal to 1 (without
loss of generality \cite{PeVeWo}), one see that for $N$ big enough
\[
\langle\psi|\psi\rangle=Tr\left(\prod_{k=1}^{N}E_{\mathbb{I}}\right)\approx\lambda_{1}^{N}=1,\]
 where $\lambda_{1}$is the maximum eigenvalue equal to $1$. The
corrections are exponentially suppressed in $N$. In our analytic
derivation of an upper bound for the Lypschtiz constant of the function
\[
f\equiv\frac{\langle\psi|\bigotimes_{k=1}^{N}S[k]|\psi\rangle}{\langle\psi|\psi\rangle}=\frac{Tr(\prod_{k=1}^{N}E_{S[k]}[k])}{Tr\left(\prod_{k=1}^{N}E_{\mathbb{I}}\right)},\]
 we assume $N$ to be big enough that for all purposes \[
f=Tr(\prod_{k=1}^{N}E_{S[k]}[k]),\]
without the need of any additional normalization (and numerically
this is verified to be true already for $N\geq4$).

\smallskip{}

\textbf{APPENDIX C.} A concentration of measure result obtained for
the trace distance between random states and their average implies
the same concentration of measure result for the expectation values.
This can be proved using the following general relation between the
trace distance of normalized states and the difference between the
expectation values of an observable $A$ \[
\vert Tr(\rho A)-Tr(\psi A)\vert=\vert Tr(\rho-\psi)A\vert\leq\Vert\rho-\psi\Vert_{1}\Vert A\Vert_{\infty},\]
 and we assume the operator norm of $A$ finite. If $\rho$ is a random
state and $\psi$ is its average, one can see that a bound on the
fluctuations of the right hand side of the above inequality implies
a bound on the fluctuations of the left hand side. In general one
can say that close states will have close expectation values for any
observable of finite operator norm. But in general the converse is
not true: if the expectation value of some observable with respect
to different states is close this does not imply that the states are
close. A way to estimate the state-distance is shown in \cite{PoShWi}.

We prove in this manuscript that for the expectation value $f$ of
any local observable, restricted to a subsystem of size $L$ much
smaller than the size $N$ of the total system, the following holds
true \[
Pr[\vert f-\overline{f}\vert>\epsilon]<c_{1}exp(-c_{2}^{\prime}\epsilon^{2}\chi/N^{2}),\]
 where the sampling is done with respect to a set of random matrix
product states of rank $\chi.$ Without lack of generality lets restrict
to a chain of qubits. Any operator in the subsystem of size $L$ can
be expressed in a basis of $4^{L}$ unitary orthogonal operators (for
an explicit construction see \cite{PoShWi}). Lets call each element
in this basis $U_{x},$ with $x\in\{1,\dots,4^{L}\}.$ We shall indicate
with $\rho$ a realization of a normalized random matrix product state
and with $\overline{\rho}$ the average state. Lets define $p_{x}(\rho)\equiv Tr(U_{x}\rho)$
and $p_{x}(\overline{\rho})\equiv Tr(U_{x}\overline{\rho})$. The
previous concentration result holds true in particular for $p_{x}$
(for any $x$)\[
Pr[\vert p_{x}(\rho)-p_{x}(\overline{\rho})\vert>\epsilon]<c_{1}exp(-c_{2}^{\prime}\epsilon^{2}\chi/N^{2}).\]
 Since there are $4^{L}$ different $x$ values we can also writhe
\[
Pr[\exists x:\vert p_{x}(\rho)-p_{x}(\overline{\rho})\vert>\epsilon]<4^{L}c_{1}exp(-c_{2}^{\prime}\epsilon^{2}\chi/N^{2}).\]
 Any random density matrix associated to the normalized RMPS can be
written as $\rho=\sum_{x}p_{x}(\rho)U_{x}$. When $\vert p_{x}(\rho)-p_{x}(\overline{\rho})\vert<\epsilon$
for all $x$, then it follows \begin{align*}
\Vert\rho-\overline{\rho}\Vert_{2}^{2} & =\Vert\sum_{x}(p_{x}(\rho)-p_{x}(\overline{\rho}))U_{x}\Vert_{2}^{2}\\
= & Tr\left[\sum_{x}\left(p_{x}(\rho)-p_{x}(\overline{\rho})\right)U_{x}\right]^{2}\\
= & 4^{L}\sum_{x}\left(p_{x}(\rho)-p_{x}(\overline{\rho})\right)^{2}<4^{2L}\epsilon^{2}.\end{align*}
 We can write then \[
\Vert\rho-\overline{\rho}\Vert_{1}\leq\sqrt{4^{L}}\Vert\rho-\overline{\rho}\Vert_{2}<4^{3L/2}\epsilon\]
 and from this it follows\[
Pr[\Vert\rho-\overline{\rho}\Vert_{1}>4^{3L/2}\epsilon]<4^{L}c_{1}exp(-c_{2}^{\prime}\epsilon^{2}\chi/N^{2}).\]
 Which tells us that keeping $L$ fixed and much smaller than $N$,
for $\chi\propto N^{3}$and $\epsilon\propto N^{-1/3}$one has\[
Pr[\Vert\rho-\overline{\rho}\Vert_{1}>4^{3L/2}N^{-1/3}]<4^{L}exp(-N^{1/3}).\]
 This prove a concentration of measure result {}``at the level of
states''. But this result is weaker with respect to the result for
the observables by a factor $4^{L}$.
\end{document}